\documentclass{article}

    \PassOptionsToPackage{ round}{natbib}

    \usepackage[preprint]{neurips_2019}

\usepackage{moreverb}
\usepackage{amsthm}
\usepackage{amsgen,amsmath,amstext,amsbsy,amsopn,amssymb}
\usepackage{mathtools}
\usepackage{graphicx}
\usepackage{caption}
\usepackage{algcompatible}
\usepackage{algorithm}
\usepackage{pdfpages}
\usepackage{graphicx}
\usepackage{flexisym}
\usepackage{subfigure}
\usepackage{color}

\newtheorem{lem}{Lemma}
\newtheorem{property}{Property}

\title{A Note on Optimal Sampling Strategy for Structural Variant Detection Using Optical Mapping}

\author{
  Weiwei Li\\
  Department of Statistics and Operations Research\\
  University of North Carolina at Chapel Hill\\
  \texttt{weiweili@live.unc.edu} \\
  \And
  Jan Hannig \\
  Department of Statistics and Operations Research\\
  University of North Carolina at Chapel Hill\\
  \texttt{jan.hannig@unc.edu} \\
  \AND
  Corbin D. Jones \\
  Department of Biology and Integrative Program for Biological $\&$ Genome Sciences\\
  University of North Carolina at Chapel Hill\\
  \texttt{cdjones@email.unc.edu} \\
}

\begin{document}

\maketitle

\begin{abstract}
  Structural variants compose the majority of human genetic variation, but are difficult to assess using current genomic sequencing technologies. Optical mapping technologies, which measure the size of chromosomal fragments between labeled markers, offer an alternative approach.  
As these technologies mature towards becoming clinical tools, there is a need to develop an approach for determining the optimal strategy for sampling biological material in order to detect a variant at some threshold.  Here we develop an optimization approach using a simple, yet realistic, model of the genomic mapping process using a hyper-geometric distribution and {probabilistic} concentration inequalities. Our approach is both computationally and analytically tractable and includes a novel approach to getting tail bounds of hyper-geometric distribution.  We show that if a genomic mapping technology can sample most of the chromosomal fragments within a sample, comparatively little biological material is needed to detect a variant at high confidence.
\end{abstract}

\section{Introduction}\label{sampling:sec:intro}
Structural variants (SV), insertions, deletions, trans-locations, copy number variants, are by far the most common types of human genetic variation \citep{chaisson15}. They have been linked to large number of heritable disorders \citep{hurles08}. Technology to assay the presence or absence of these variants has steadily improved in ease and resolution \citep{huddleston16,audano19}. Whole genome shotgun DNA sequencing (WGS) can detect small variants (less than 10bp) readily and can detect some classes of large SV. This approach, however, is inferential and often struggles to capture copy number variation in gene families or to correctly estimate the size of insertions. An alternative approach, genomic mapping (such as the technology of BioNano Genomics), addresses the deficiencies of WGS by providing linkage and size information from ordered fragments of chromosomes spanning tens to hundreds of kilobases. In contrast to WGS, genomic mapping approaches directly observe SV, rather than inferring the existence of a SV from patterns of mismatch in WGS data.  In the near future, these genome mapping technologies are expected to be used for clinical diagnosis of SV known to be associated with genetic disorders. 

In a clinical setting, the cells or tissues needed for analysis may be hard to obtain, which poses several important statistical questions:  what is the minimum amount of starting material necessary to have some confidence of detecting a target fragment? What is the optimal sampling strategy for the primary and derived material throughout the process?  How best to model the technical errors--such as failure to digest at a site--during the processing of the data as these errors can lead to false positives and negatives?   As is often the case, answering these questions motivated an exploration and expansion of the statistical machinery used to model this biological process. Specifically, we established a relationship between the tail bounds of the binomial and hyper-geometric distributions.

\section{Statistical Model}\label{sampling:sec:model}
\begin{figure*}[b]
\captionof{figure}{Urn demonstration of sampling procedure}\label{sampling:fig:01}
\centerline{\includegraphics[width=1\textwidth]{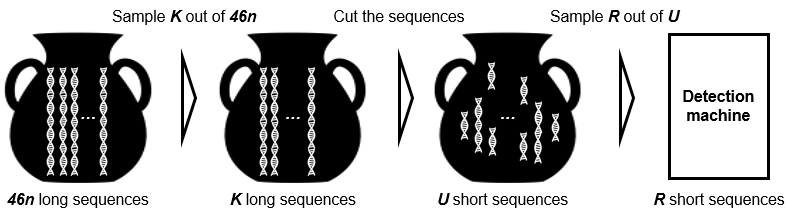}}
Three urn demonstration of the algorithm. The first urn contains raw biological materials. The second urn contains materials sampled from the first urn. The third urn contains materials from the second urn that are cut into shorter segments. Content of the third urn is sampled and assayed in the detection machine.
\end{figure*}

In this section, we abstract our sampling procedure into an ``urn sampling" model.  As DNA is processed through the optical mapping procedure, we imagine the material passing through a series of urns.  Assume we have $46$ different types of long sequences (i.e. chromosomes), each type has $n$ copies (i.e. $n$ cells), so we have $46 n$ long sequences in total. We assume only one type of long sequences contains the target sequence, or the fragment of interest.  
{The basic idea of our sampling model is shown in Figure~1\vphantom{\ref{sampling:fig:01}}. 
The notations introduced below are summarized in Table~\ref{sampling:Tab:01}.}

\begin{table}[t]
\captionof{table}{Nonrandom Quantities}\label{sampling:Tab:01}
\centering
\begin{tabular}{@{}cl@{}}
\hline
\textbf{Notation} & \multicolumn{1}{c}{\textbf{Definition}}                              \\ \hline
n                 & Number of cells in the first urn (copies of each type of long sequences).   \\
K                 & Number of sequences sampled from the first urn.                      \\
R                 & Number of sequences sampled from third urn.                          \\
L                 & Approximated length of long sequence.                                \\
l                 & Approximated length of short sequence.                               \\
T                 & Threshold on detectability of target sequences.                      \\
f                 & Length of fragment of interest.                                      \\
c                 & Approximated ratio between lengths of long and short sequences.      \\
Q                 & Minimum number of target sequences we want in the detection machine. \\
p                 & Minimum confidence in achieving the goal.                            \\ \hline
\end{tabular}
\end{table}

{
The first urn contains our original biological material, total of $46 n$ long sequences out of which $n$ of them contain the target sequence. At the first stage, we sample $K$ sequences without replacement from the first urn, and put them in the second urn. The second urn will therefore contain a random number $X$ of target sequences. All of the $K$ long sequences in second urn are cut at random locations according to a Poisson process and placed into the third urn. The third urn will therefore contain a random number of $U$ sequences out of which $W$ are target sequences. The content of the third urn models the biological material prepared for assay in a detection machine. Finally, we sample $R$ smaller sequences without replacement out of the third urn and put them into a detection machine. There will be a random number $Y$ of target sequences processed by the detection machine, and the goal is to assure that for some pre-specified values $Q$ and $p$, we have the probability of $Y\geq Q$ is at least $p$. Throughout the experiment, the variables ($n, K, R$) are in our control and we will find the conditions on them to achieve our goal. Throughout this paper, we call the long sequence in the second urn which contains the fragment of interest as ``target sequence''.}

Next we state the following biological assumptions:
\begin{enumerate}
\item The length of target sequence is $f$.
\item The lengths of long sequences in the first urn are approximately $L$, here $L \gg max(f,T)$.
\item Short sequences in the third urn have lengths approximately  $l$, and we have $c \approx \frac{L}{l}$.
\end{enumerate}

We proceed by describing the probabilistic parts of our model. The distributions and their expectations are summarized in Table~\ref{sampling:Tab:02}. There are $X$ target sequences in the second urn. It is straightforward to see $X \sim H(46n,n,K)$, a hyper-geometric distribution with $46n$ samples and $n$ samples of interest and $K$ as sampling size. Hence $\mathbb{E}[X] = \frac{K}{46}$. 

 Let $U_i$ ($i=1,2,..,K$) denotes the number of cuts on $i$-th long sequence in the second urn. Combine with the third assumption above, we assume that $U_i$ follows a Poisson distribution with mean $c$. Note that $U_i$ cuts divide the sequence into $(U_i+1)$ shorter sub-sequences. Consequently,  $U=\sum_{i=1}^{K}(U_i+1)$ is the total number of short sequences in the third urn, and $(U-K)$ follows Poisson distribution with mean $cK$.

{Write $W$ as the number of the sequences in the third urn that contain the target sequence. The distribution of $W$ is more complicated than that of $X$. Assuming $X>0$,  we have at least $1$ target sequence contained in the second urn. We have $W=\sum_{i=1}^{X} B_i$, where fix $X$, $\{ B_i \}_{i=1}^X$ are independent Bernoulli random variables.
Condition on $\{ U_i \}_{i=1}^{K}$, the probability of success $q_i$ of random variable $B_i$ satisfies 
\begin{equation}\label{sampling:eq:ber_probability}
   q_i(U_i) \begin{cases*}
       \geq 2(t_1t_3)^{U_i}-(t_2t_3)^{U_i} & if $T\geq f$, \\
      =t_3^{U_i}       & otherwise,
    \end{cases*}
  \end{equation}
respectively. Here $t_1=\frac{L-T}{L-f}$, $t_2=\frac{L-2T+f}{L-f}$, $t_3=1-\frac{f}{L}$.}
The proof is found in Section~\ref{sampling:sec:bernoulli}.

Finally, condition on $U$ and $W$, the number of target sequences in the detection machine $Y$ follows a hyper-geometric distribution with parameters $U$, $W$ and $R$.

\begin{table}[t]
\captionof{table}{Random Quantities and Their Expectations}\label{sampling:Tab:02}
\centering
\begin{tabular}{@{}ccc@{}}
\hline
\textbf{Notation} & \textbf{Distribution}        & \textbf{Expectation}                        \\ \hline
X                 & $H(46n,n,K)$                 & $\frac{K}{46}$                              \\
$U_i$             & $Poi(c)$                     & $c$                                         \\
$W$               & $\sum_{i=1}^X Ber(q_i(U_i))$ & $\frac{K(2e^{ct_1t_3}-e^{ct_2t_3})}{46e^c}$ \\
$Y \mid U, W$     & $H(U,W,R)$                   & $\frac{WR}{U}$                              \\ \hline
\end{tabular}
\end{table}

\subsection{Analytical Results}\label{sampling:sec:analytical}
In this section, we present the analytical results of our statistical modeling. Mathematically, our goal can be written as 
\begin{equation}\label{sampling:eq:goal}
P(Y \geq Q) \geq p,\textit{ for pre-specified $Q$ and $p$.}
\end{equation}

Now we consider the quantity $R_{low}$, such that with pre-fixed quantities $p_0$, $U$ and $W$ 
\begin{equation} \label{sampling:eq:Rbound}
 P(Y\geq Q\mid U,W, R \geq R_{low}) \geq p_0.
\end{equation}
Note here $Y | U, W \sim H(U,W,R)$. We will find $R_{low}$ as a function of $U, W, p_0$ from a concentration inequality on hyper-geometric distribution. 

In section~\ref{sampling:sec:hyper_proof}, we developed the relationship between the tail bounds of binomial distribution and that of hyper-geometric distribution. Specifically, consider the following random variables with parameters $A, B, C$:
\begin{enumerate}
\item $h \sim H(A,B,C)$, a hyper-geometric distributed random variable.
\item $B_a \sim Bin(C,\frac{B}{A})$ and $B_b \sim Bin(A-C,\frac{B}{A})$.
\end{enumerate}
We proved that under some regularity conditions, the following relations are true  
\begin{align}
&P(h \leq x) \leq P(B_a \leq x), \label{sampling:eq:goal_1}\\
&P(h \leq x) \leq P(B_b \leq B-x), \label{sampling:eq:goal_2}
\end{align}
The conditions needed and detailed proof are presented in section~\ref{sampling:sec:hyper_proof}. 

In section~\ref{sampling:sec:numerical}, the numerical calculations implied that for large $C$, \eqref{sampling:eq:goal_2} is a better bound, otherwise we may want to use \eqref{sampling:eq:goal_1}. From the relationship above we immediately know a tail bound on binomial distribution can also be used as the tail bound for hyper-geometric distribution. 

Throughout this paper, we assume the conditions needed for \eqref{sampling:eq:goal_1} and \eqref{sampling:eq:goal_2} are always met. Therefore we may use large deviation bounds from \cite{arratia89} at the following two binomial distributions: $Bin(R, \frac{W}{U})$ and $Bin(U-R, \frac{W}{U})$ to find $R_{low}$ in \eqref{sampling:eq:Rbound}. 

From now on we write $R_{low} = R_{low}(U,W,p_0)$. Note that $U$ and $W$ are typically unknown. Therefore, $R_{low}$ itself is still a random quantity and we need to further find a upper bound for $R_{low}$ depending on $n$ and $K$, this is denoted by $\hat{R}_{low}$. With large probability, sampling $\hat{R}_{low}$ sequences in the third urn is enough to guarantee sampling no less than $R_{low}$ samples. 

It is fairly straightforward to see $R_{low}$ increases with $W$ and decreases with $U$. Now we fix $Q$ and $p_0$, and write $U_{up}$ and $W_{low}$ as the probabilistic upper/lower bounds for $U$ and $W$, respectively. From \eqref{sampling:eq:goal_1} and \eqref{sampling:eq:goal_2} we can find $\hat{R}_{low}$ directly from tail bounds on $Bin(R, \frac{W_{low}}{U_{up}})$ and $Bin(U_{up}-R, \frac{W_{low}}{U_{up}})$. In particular, the steps
needed to determine $\hat{R}_{low}$ for a given $K$ and $n$ are summarized here:
\begin{enumerate}
\item Use lemma~\ref{sampling:lem:largedeviation} on binomial distributions $Bin(K, \frac{1}{46})$ and $Bin(46n-K, \frac{1}{46})$ to find lower bound $X_{low}$ of $X$. Here $X_{low}$ depends only on $n$, $K$ and $p_1$ so that: $P(X \geq X_{low}) \geq p_1$.

\item Set $X:=X_{low}$ from step 1. Note that $W$ is the summation of $X_{low}$ independent Bernoulli trials. Hence from lemma~\ref{sampling:lem:largedeviation} we can find lower bound $W_{low}$ of $W$ depending only on $n$, $K$, $L$, $f$, $T$, $c$, $p_1,p_2$ so that:$P(W \geq W_{low} \mid X\geq X_{low}) \geq p_2.$ Consequently $P(W \geq W_{low}) \geq p_1 p_2 $.

\item Use inequality from lemma~\ref{sampling:lem:poisson} to find $U_{up}$  and $U_{low}$ depending only on $c,K,p_3$ so that: $P(U \geq U_{low}) \geq p_3$ and  $ P(U \leq U_{up}) \geq p_3$.

\item Use lemma~\ref{sampling:lem:largedeviation} on binomial distributions $Bin(R, \frac{W_{low}}{U_{up}})$ and $Bin(U_{up}-R,\frac{W_{low}}{U_{up}})$ to find $\hat{R}_{low}$ so that:
\begin{align*}
P(\hat{R}_{low} \geq R_{low}) & \geq P(U \leq U_{up}, W \geq W_{low}) \\
& \geq P(U \leq U_{up})+P(W \geq W_{low})-1 \\
&=p_3+p_1p_2-1.
\end{align*}
\end{enumerate}

Note that we need to ensure the needed sample size $R$ is not larger than the available number of short sequences $U$. To this end, both $\hat{R}_{low}$ and $U_{low}$ are deterministic functions of given constants and we can add numerical constraint on $\hat{R}_{low}$ to force it smaller than $U_{low}$. A key observation from our numerical result is, as $K$ gets larger, $U_{up}$ and $U_{low}$ will be more concentrated around the mean $cK+K$, while $R_{low}$ will be much smaller than $U_{low}$. Therefore, we need to find a lower bound $K_{min}$ on $K$ to ensure $U_{low} \geq \hat{R}_{low}$.

Finally, given that we choose $K$ and $\hat{R}_{low}$ as our sampling sizes at two stages, respectively. The following relations are true:
\begin{align}
P(Y \geq Q) & \geq P(Y \geq Q, R \geq R_{low}, U \geq R) \nonumber \\
& \geq p_0 \cdot P(\hat{R}_{low} \geq R_{low}, U \geq \hat{R}_{low}) \nonumber \\
& \geq p_0 \cdot \left [ P(\hat{R}_{low} \geq R_{low})+P(U \geq \hat{R}_{low})-1 \right] \nonumber \\
&\geq p_0 (2p_3+p_1 p_2-2). \label{sampling:eq:calp}
\end{align}
It suffices to set the desired probability $p$ equal to the right-hand-side of \eqref{sampling:eq:calp}. The exact 
selection of $\{ p_i \}_{i=0}^{3}$ can be found in Section~\ref{sampling:sec:algo}. As discussed in section~\ref{sampling:sec:algo}, the range of $K$ is $[K_{min}, 45n]$, while not every $K$ in this range is feasible, a straightforward monotone analysis shows that as long as $K$ is larger than a certain threshold, the solution $\hat{R}_{low}$ always exists.

\subsection{Optimal Sampling Strategy} \label{sampling:subsec:logic}
In this section, we discuss how to use the formulas derived in section~\ref{sampling:sec:analytical} to find the optimal values of $n$ and $K$ for any given $p$ and $Q$. Specifically, assume there is a user-specified cost function $f(n,K)$ over number of samples $n$ and the sampling size from first urn. In this paper we assume $f(\cdot,\cdot)$ is an monotone increasing function of both $n$ and $K$.
 
The proposed procedure is summarized here:
\begin{enumerate}
\item Solve for $\{ p_i \}_{i=0}^3 $ such that $p = p_0(2p_3 + p_1 p_2-2)$.
\item For fixed $n$, we calculate $K_{min}$.
\item For any fixed $n$ and $K$ such that $K \geq K_{min}$, we calculate $\hat{R}_{low}$.
\item Return: $(n,K,\hat{R}_{low})$.
\end{enumerate}

The implementation details are discussed in section~\ref{sampling:sec:appen}. In reality the amount of biological materials is limited, hence there is an upper bound on $n$ and there are only finite number of $(n,K,\hat{R}_{low})$ to consider. We do not need to consider any $R>\hat{R}_{low}$ as that would lead to sub-optimal design. However, for fixed $n$, we do need to consider $K > K_{min}$, because larger $K$ might lead to smaller $\hat{R}_{low}$ and a more efficient solution. 

Assume we have a cost function $C(K,R)$ that increases with $K$ and $R$. We only have finitely many $(n,K,\hat{R}_{low})$ to consider and a brute force search among all the possible triples will yield the optimal $(n,K,\hat{R}_{low})$ minimizing the cost function. 

Due to technology limits, we may have certain constraints on sampling percentages: for example, we can only sample $80\%$ in the first stage, and $50\%$ from the second stage. We can still use the brute force search only considering the cases that do satisfy these extra constraints.

\section{Numerical Results and Conclusions}\label{sampling:sec:numerical}

For our numerical results, the calculations were based on biologically reasonable parameters: $L=250000000$, $f=50000$, $T=75000$, $c=60$, $p=0.95$, $Q=20$.

\begin{figure*}
\centering
\captionof{figure}{Results on approximation}\label{sampling:fig:02}
\centerline{\includegraphics[width=8cm,height=8cm]{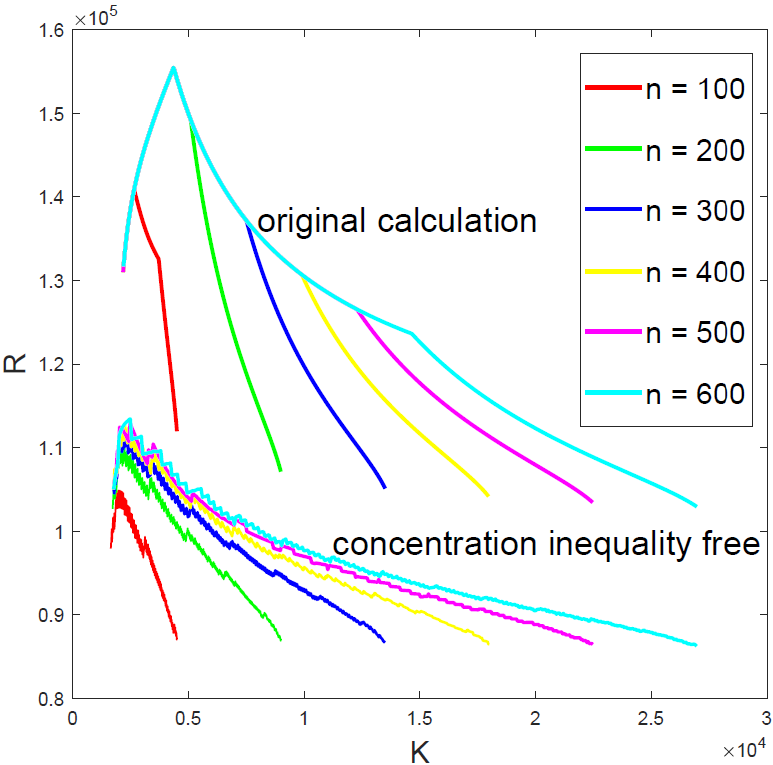}}
Plot of $K$ vs, $\hat{R}_{low}$ using Algorithm~\ref{sampling:Algo:01} for $n$ ranges from $n=100$  to $n=600$, different colors correspond to different $n$. Curves at the bottom correspond to concentration inequality free results. 
\end{figure*}

\begin{figure*}
\captionof{figure}{Results on simulation}\label{sampling:fig:03}
\centerline{\includegraphics[width=8cm,height=8cm]{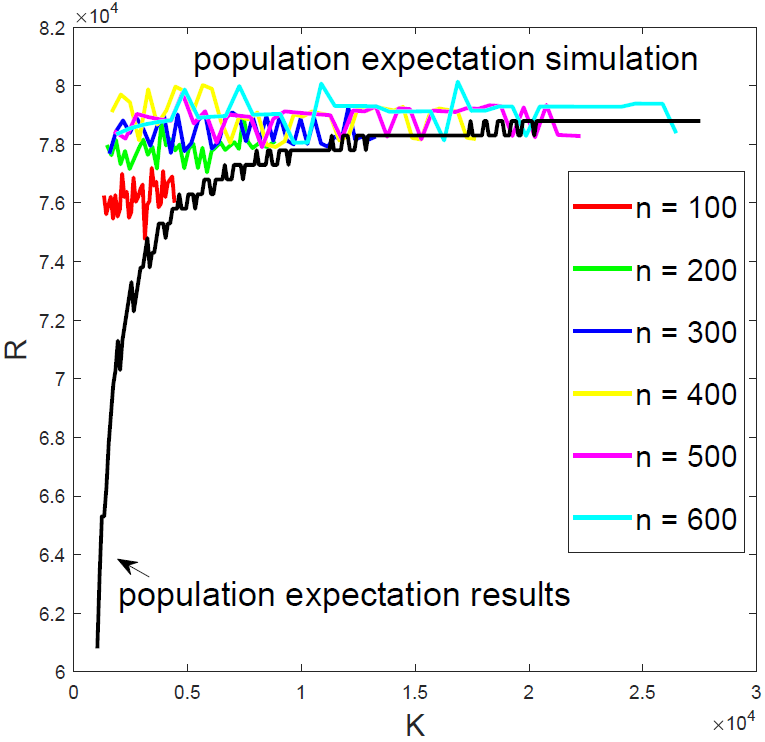}}
Plot of simulation results and population expectation results, here $n$ ranges from $100$ to $600$.
\end{figure*}

In Figure~\ref{sampling:fig:02}, we plot our original calculation results from Algorithm~\ref{sampling:Algo:01} together with the results without using any concentration inequalities (we get the tail points by the inverse of cumulative distribution functions, which is applicable for relatively small $n$); both of them have the similar patterns. From original calculation results we can find two ``kinks'' for each fixed $n$. This is because when $K$ is small, we will need to sample almost everything from the second stage, which will force us to choose the correspond $Bin(U_{up}-R, \frac{W_{low}}{U_{up}})$ for $Y$ as the binomial bounds. Then as $K$ gets larger but not big enough, we will use $Bin(R, \frac{W_{low}}{U_{up}})$ for both stages. Finally $K$ will get close to $45n$ which again forces to use $Bin(U_{up}-R, \frac{W_{low}}{U_{up}})$ at the first sampling stage. 

In Figure~\ref{sampling:fig:03} we plot the simulation results together with population expectation results. Here the simulation means of each $n$ and fixed $K$ we create large amount of $X$, $W$ and $U$. Then for each simulation trial, we use a brute force search to find the smallest $R$ that can gives us \eqref{sampling:eq:goal}. Note this simulation is an ``averaging'' approach while our algorithm is more like a tolerance interval approach, thus they are not comparable and we put them into two separate figures. The population expectation results means we replace $W$ and $U$ directly by their expectations, and again brute force search for the smallest $R$. From Figure~\ref{sampling:fig:03} we can see as $K$ gets larger, these two results will be very close, which implies for large $K$, we can approximately use expectations of $U$ and $W$ to conduct the calculation. 

Table~\ref{sampling:Tab:03}, provides examples the minimization results based on a linear cost function.
$
C(K,R)=a  K+b R
$
under various constraints. 
In particular we use $a=60$, $b=1$ and various sampling percentage constraints on both sampling stages.
\begin{table}[t]
\captionof{table}{Minimization of Cost Function}\label{sampling:Tab:03}
\centering
\begin{tabular}{@{}ccccccc@{}}\hline
Constraint $1$ & Constraint $2$ & $n$ & $K$ & $R$ & $\frac{K}{46n}$ & $\frac{R}{U_{low}}$ \\ \hline
$100\%$ & $50\%$ & $100$ & $4048$ & $123696$ &$97.71\%$ & $49.86\%$\\
$80\%$ & $20\%$ & $300$ & $9868$ & $120365$ & $71.51\%$ & $19.93\%$ \\
$50\%$ & $100\%$ & $500$ & $2168$ & $131001$ &$9.43\%$ & $98.41\%$ \\
$50\%$ & $50\%$ & $500$ & $4918$ & $150542$ &$21.38\%$ & $49.96\%$ \\
$20\%$ & $80\%$ & $600$ & $2968$ & $144549$ & $10.75\%$ & $79.4\%$ \\ \hline
\end{tabular}
\end{table}
 
We have also applied our algorithm to other choices of $Q$. The lessons learned are similar to what we have shown here. In the supporting materials we provide the Matlab code that can be used to calculate optimal sampling strategy with different parameters.

In conclusion, we have developed an optimization approach for estimating the amount of material needed for genomic mapping based on a simple, yet realistic, model of the process that uses a novel result regarding the tail bounds of the hyper-geometric distribution. Our approach is both computationally and analytically tractable and We show that if a genomic mapping technology can sample most of the chromosomal fragments within a sample, comparatively little biological material is needed to detect a variant at high confidence.

\section{Appendix}\label{sampling:sec:appen}
\subsection{Proof of Equation \eqref{sampling:eq:ber_probability}}\label{sampling:sec:bernoulli}
\begin{proof}
There are $X$ copies of the target fragments in the second urn. Some of the fragments of interest might not survive during the cutting process, therefore we have $W\leq X$. Define $\{ A_i \}_{i=1}^X $ as the event that the $i$-th target fragment survives (i.e. being intact after cutting procedure) and is placed in the third urn. Given that we have $U_i$ cuts on the $i$-th target sequence, the locations of these $U_i$ cuts are then uniformly distributed, therefore $p(A_i)=(1-f/L)^{U_i}$. 

Next, in order for the target fragment to be usable by the detector, it has to be longer than $T$.  If $T \leq f$, the sequences that contain the target fragment are always longer than $T$, then $q_i(U_i)=p(A_i)$. Otherwise we estimate $q_i$ from a lower bound using the probability of an event $A_i\cup E_i$, where $E_i$ is the event of not having cuts within $T-f$  on either one or the other side of the target sequence (see Figure~\ref{sampling:fig:04}). Recall that $t_1=\frac{L-T}{L-f}$, $t_2=\frac{L-2T+f}{L-f}$, $t_3=1-\frac{f}{L}$.
Then by inclusion and exclusion  $ p(E_i \mid A_i)$= $2(t_1)^{U_i}-(t_2)^{U_i}$ and consequently 
\[
q_i(U_i) \geq p(A_i)p(E_i \mid A_i)=2(t_1t_3)^{U_i}-(t_2t_3)^{U_i}.
\]

\begin{figure*}[h]
\captionof{figure}{Demonstration of cutting}\label{sampling:fig:04}
\centerline{\includegraphics[width=0.48\textwidth]{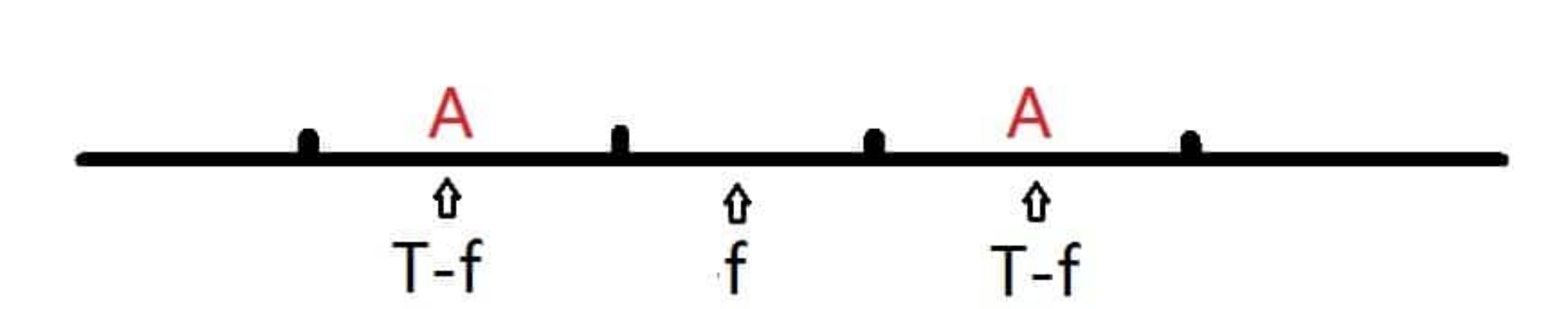}}
DNA sequence with target fragment. The $f$ zone and at least one of the $A$ zones should have no cuts to provide a valid target sequence.
\end{figure*}

\end{proof}

\subsection{Hyper-geometric Distribution and Binomial Bounds} \label{sampling:sec:hyper_proof}
In this section, we discuss the relationship between the tail bounds of binomial distribution and that of hyper-geometric distribution. 

For fixed positive integer $x$, consider the following two inequalities
\begin{align} 
&P(H = x) \leq P(B_1 = x) \label{sampling:eq:goal_1_pt1},\\
&P(H = x) \leq P(B_2 = B-x) \label{sampling:eq:goal_2_pt1},
\end{align}
The above two inequalities can be simplified as:
\begin{equation}\label{sampling:eq:sim_1}
\frac{ \binom{A-C}{B-x}}{\binom{A}{B}} \leq r^x(1-r)^{C-x},
\end{equation}
and 
\begin{equation}\label{sampling:eq:sim_2}
\frac{ \binom{C}{x}}{\binom{a}{b}} \leq r^{B-x}(1-r)^{A-B-C+x},
\end{equation}
respectively. Note that if \eqref{sampling:eq:goal_1_pt1} is true for all $x' \leq x_0$, then \eqref{sampling:eq:goal_1} is true for $x = x_0$, similarly for $\eqref{sampling:eq:goal_2_pt1}$. Now we discuss the following properties for \eqref{sampling:eq:sim_1} and \eqref{sampling:eq:sim_2}.
\begin{property}\label{sampling:pro:01}
\emph{For fixed $B$, $A$, $C_0$, and $x \leq \frac{BC_0}{A}$, if \eqref{sampling:eq:sim_1} and \eqref{sampling:eq:sim_2} are true for $C=C_0$, then they are also true for any $C$ such that $C_0 \leq C < A-B$.
}
\end{property}
\begin{proof}
We use mathematical induction on $C$. Given that $\frac{ \binom{A-C_0}{B-x}}{\binom{A}{B}} \leq r^x(1-r)^{C_0-x}$. Now for $C:=C_0+1$, we want $\frac{ \binom{A-C_0-1}{B-x}}{\binom{A}{B}} \leq r^x(1-r)^{C_0+1-x}$. It suffices to have $\frac{ \binom{A-C_0-1}{B-x}}{\binom{A}{B}} \leq \frac{ \binom{A-C_0}{B-x}}{\binom{A}{B}}(1-r)$, which only requires $x \leq \frac{BC_0}{A}$. Similarly we can prove this property for \eqref{sampling:eq:sim_1}.
\end{proof}

\begin{property}\label{sampling:pro:02}
\emph{
For fixed $B$, $A$ and $C$. If \eqref{sampling:eq:sim_1} and \eqref{sampling:eq:sim_2} are true for some fixed $x=k \leq \frac{BC}{A}$, then they are also true for any $x$ such that $x \leq k$.
}
\end{property}
Proof of Property~\ref{sampling:pro:02} is almost the same as that of Property~\ref{sampling:pro:01}, hence we omit it here.

\begin{property}\label{sampling:pro:03}
Assume the following inequalities are true for some constants $A_{up}$ and $B_{low}$
\begin{align*}
    2A \geq 2B+C, \quad Q \leq g \frac{BC}{A},\quad A \geq \frac{3}{(3-2g)}B+\frac{2}{(3-2g)}C+\frac{3}{(3-2g)}\frac{3A}{C}, \\
    x \geq 5, \quad \Phi(A_{up}) \geq 0, \quad (B-Q)C \geq B(2Q+1), \quad A \geq 3B+C
    G(B_{low}) \geq 0,
\end{align*}
where $g = \frac{xA}{BC}$ and 
\begin{align*}
     \Phi(A)=&(C-1)\log(A)+(C-x)\log(A-B-1)-C \log(A-1)-(C-x-1)\log(A-B) \\
     & +\log(A-C)-\log(A-B-C+x), \\
     G(B)=&(x-1)\log(B+1)+(C-x)\log(A-B-1) -x \log B-(C-x-1)\log(A-B) \\
     & +\log(1+B-x)-\log(A-B-C+x).
\end{align*}
Then for fixed $B$, $x$ and $C$, if \eqref{sampling:eq:sim_1} is true for $A=A_{up}$, it is also true for $A \leq A_{up}$; for fixed $A$, $x$ and $C$, if \eqref{sampling:eq:sim_1} is true for $B=B_{low}$, then it is also true for $B \geq B_{low}$.
\end{property}
\begin{proof}
Again we use (backward) mathematical induction on $A$. Given $\frac{ \binom{A-C}{B-x}}{\binom{A}{B}} \leq r^x(1-r)^{C-x}$. We need $
 \frac{ \binom{A-1-C}{B-x}}{\binom{A-1}{B}} \leq (\frac{B}{A-1})^x(1-\frac{B}{A-1})^{C-x}$. It suffices to show 
\begin{align*}
\frac{ \binom{A-1-C}{B-x}}{\binom{A-1}{B}} \leq(\frac{B}{A-1})^x(1-\frac{B}{A-1})^{C-x} (\frac{A}{B})^x(1-\frac{B}{A})^{x-C} \frac{ \binom{A-C}{B-x}}{\binom{A}{B}},
\end{align*}
the inequality above is equivalent to $\Phi(A) \geq 0$. Take first order derivative of $\Phi(A)$ with respect to $A$ we have:
\begin{equation*}
\begin{split}
\Phi'(A)=-\frac{C}{A-1}+\frac{C-1}{A}+\frac{1}{A-C} -\frac{C-x-1}{A-B}+\frac{C-x}{A-B-1}-\frac{1}{A-B-C+x}.
\end{split}
\end{equation*}
If $\Phi'(A) \leq 0$ for $A \leq A_{up}$, the result is proved by using the monotonicity of $\Phi(A)$ and the assumption that $\Phi(A_{up}) \geq 0$. Now we will prove $\Phi'(A) \leq 0$. It suffices to show: 
\begin{equation}
\label{sampling:eq:01} 
\begin{aligned}
& -B^3C^2+B^3 C-B^2 C^3+B^2 C^2 x-B^2 C x +B^2 C-B C^3+B C^2 x+B C^2-B C x  \\
& +A (3 B^2 C^2-3 B^2 C+2 B C^3-2 B C^2 x+2 B C x-2 B C +C^2 x-C x^2)  \\
& +A^2(-3 B C^2+3 B C-C^2 x+C x^2-2 C x+x^2+x) +A^3 (2 C x-x^2-x) \leq 0.
\end{aligned}
\end{equation}
The fist line of \eqref{sampling:eq:01} is obviously negative by noting the following facts
\begin{align*}
& B^2C^2x \leq B^2C^3, \quad  B^2C \leq B^2Cx, \quad B^3C \leq B^3C^2, \quad BC^2x+BC^2 \leq BC^3+BCx.
\end{align*}
For the rest lines, we use the following relations:
\begin{align*}
 Ax^2+Ax+ACx^2 \leq 3AB^2C+2BC^2x, \quad C^2x+2BCx \leq 2ACx, \quad -x^2-x \leq 0,
\end{align*}
where the last inequality follows by assumption $2A \geq 2B+C$. For the rest parts, we want $2xA^2+3AB+2BC^2+3B^2C \leq 3ABC$. It suffices to show
\begin{align*}
(3-2g)A \geq 3B+2C+\frac{3A}{C}, 
\end{align*}
which is equivalent to 
\begin{align*}
 A \geq \frac{3}{(3-2g)}B+\frac{2}{(3-2g)}C+\frac{3}{(3-g)}\frac{3A}{C},
\end{align*}
this follows directly from the assumptions. Thus the first part of Property $3$ is proved.

Now we prove the second part. From mathematical reduction on $B$, we want $\frac{ \binom{A-C}{B+1-x}}{\binom{A}{B+1}} \leq (\frac{B+1}{A})^x(1-\frac{B+1}{A})^{C-x}$. It suffices to have
\begin{align*}
\frac{ \binom{A-C}{B+1-x}}{\binom{A}{B+1}} \leq& (\frac{B+1}{A})^x(1-\frac{B+1}{A})^{C-x} \cdot (\frac{A}{B})^x(1-\frac{B}{A})^{x-C}\frac{ \binom{A-C}{B-x}}{\binom{A}{B}}, 
\end{align*}
which is equivalent to $G(B) \geq 0$. Similarly as before, we want this function increases with $B \geq B_{low}$, from which we only need to check $G(B_{low}) \geq 0$ and this follows from our assumption. Consider the first order derivative of $G(B)$:
\begin{equation*}
G'(B)= \frac{1}{1+B-x}+\frac{C-x-1}{A-B}-\frac{C-x}{A-B-1}+\frac{x-1}{1+B}-\frac{x}{B}+\frac{1}{A-B-C+x}.
\end{equation*} 
Then $G'(B)\leq 0$ requires 
\begin{equation}
\label{sampling:eq:02}
\begin{aligned}
& -B^3(C-1)(C-2x)+B^2C^2(x-2)-BC(x-1) + BC^2(x-1)-3B^2x + B(x-1)x \\
& -BC(x-1)x+B^2x^2+B^2C(2+2x-x^2) + A^3(1-x)x\\
&  +A^2\left[ (x-1)x+3B(x-1)x+C(x-1)x+(1-x)x^2 \right ] \\
& +A(-3B^2(x-1)x-C(x-1)x-2BC(x-1)x +2B(x-1)^2x+(x-1)x^2) \leq 0.
\end{aligned}
\end{equation}
 We can expand the first line of \eqref{sampling:eq:02} and write it as:
\begin{align*} 
&-B^3C^2+B^3(2x+1)C-2xB^3+B^2C^2x-2B^2C^2-BCx+BC+BC^2x-BC^2-3B^2x \\
&+Bx^2-Bx-BCx^2+BCx+B^2x^2,
\end{align*}
we want to show the above line is non-positive. Note that 
\begin{align*}
&-BCx+BC \leq 0, \quad BC^2x-2B^2C^2 \leq 0, \quad -BC^2 \leq 0\\
&-Bx \leq 0, \quad -3B^2x+Bx^2 \leq 0, \quad -BCx^2+BCx \leq 0, \quad B^2x^2-2xB^3 \leq 0. 
\end{align*}
Finally we only need $-B^3C^2+B^3(2x+1)C+B^2C^2x \leq 0$, which follows from our assumption: $(B-x)C \geq B(2x+1)$. 

From $x \geq 5$ we immediately get: $2+2x-x^2 \leq 0$, hence $B^2C(2+2x-x^2)\leq0$. For the second and third terms at the second line of \eqref{sampling:eq:02} we show:
\begin{align*}
 A(1-x)x+(x-1)x+3B(x-1)x+C(x-1)x+(1-x)x^2 \leq 0,
\end{align*}
it suffices to have $A-3B-C \geq 0$, which is our assumption. It is fairly straightforward to prove the last line of \eqref{sampling:eq:02} is non-negative, hence we omit it here.
\end{proof}

\begin{property}\label{sampling:pro:04}
Assume the following inequalities are true for constants $A_{up}$ and $B_{low}$
\begin{align*}
    \Phi(A_{up}) \geq 0, \quad Ax \geq B+x+2Bx, \quad G(B_{low}) \geq 0,
\end{align*}
where 
\begin{align*}
     \Phi(A)=&(A-B-C+x-1)\log(A-1-B)+(A-C)log(A)-(A-C-1)log(A-1) \\
    & -(A-B-C+x)\log(A-B)+\log(A-B)-\log(A),  \\
     G(B)=&(B-x)\log(B+1)+(A-B-C-1+x)log(A-1-B)-(B-x)\log(B) \\   &-(A-B-C+x-1)\log(A-B).
\end{align*}
Then for fixed $B$, $x$ and $C$, if \eqref{sampling:eq:sim_2} is true for $A=A_{up}$, it is also true for $A \leq A_{up}$; for fixed $A$, $x$ and $C$, if \eqref{sampling:eq:sim_2} is true for $B=B_{low}$, then it is also true for $B \geq B_{low}$.
\end{property}
\begin{proof}
Same as before we use (backward) mathematical induction on $A$. For $A = A_{up}$ we want:
\begin{align}
& \frac{ \binom{C}{x}}{\binom{A-1}{B}} \leq \frac{ \binom{C}{x}}{\binom{A}{B}} (\frac{B}{A-1})^{B-x}(\frac{A-1-B}{A-1})^{A-B-C+x-1} r^{x-B}(1-r)^{-A+B+C-x}, \nonumber 
\end{align}
which is equivalent to $\Phi(A_{up}) \geq 0$. Similarly as the proof of Property $3$, it suffices to show 
\begin{align*}
& \Phi'(A)=\frac{C+1-A}{A-1}+\frac{A-C-1}{A}+\frac{A-B-C+x-1}{A-B-1}-\frac{A-B-C+x-1}{A-B} + \log \frac{A(A-B-1)}{(A-1)(A-B)} \leq 0,
\end{align*}
for any $A \leq A_{up}$ that satisfies the assumptions. It suffices to have $B + B^2 + BC + B^2 C + A (-B - 2 BC - x) + A^2 x \leq 0 $, this only needs $ B+1 \leq A$, which is obviously true according to our assumptions. Similarly for the second part we need for $B=B_{low}$:
\begin{align*}
\frac{ \binom{C}{x}}{\binom{A}{B+1}} \leq \frac{ \binom{C}{x}}{\binom{A}{B}} (\frac{B+1}{A})^{B+1-x}(\frac{A-1-B}{A})^{A-B-C+x-1} r^{x-B}(1-r)^{-A+B+C-x},
\end{align*}
and it suffices to have $G(B) \geq 0$ for any $B \geq B_{low}$ that satisfies the assumptions. Again we prove the monotonicity of $G(B)$:
\begin{align*}
G'(B)= &\frac{-B-C+A+x-1}{A-B}-\frac{-B-C+A+x-1}{-B+A-1}-\log (-B+A-1)+\log (A-B)  \\
& -\frac{B-x}{B}+\frac{B-x}{B+1}-\log (B) +\log (B+1) \geq 0,
\end{align*}
it suffices to show $B + B^2 + BC + B^2 C - BA - A x - 2 B A x + A^2 x \geq 0$, which can be proved by using our assumption $Ax \geq B+x+2Bx$. Thus the second part is proved.
\end{proof}

\subsection{Implementation Details} \label{sampling:sec:algo}
In this section we discuss the implementation details of optimal sampling strategy in section~\ref{sampling:sec:model}.

The following quantities should be specified/calculated beforehand:
\begin{enumerate}
\item Specify the values of $L$, $f$, $T$, $p$, $Q$, $n$, $c$ according to the particular application.
\item Select $p_0=\sqrt{p}$, $3p_3-2=\sqrt{p}$ and $p_1=p_2:=\sqrt{p_3}$ so that the right-hand-side of \eqref{sampling:eq:calp} becomes $p$.
\vspace{0.1cm}
\item Compute: $t_1=\frac{L-T}{L-f}$, $t_2=\frac{L-2T+f}{L-f}$, $t_3=1-\frac{f}{L}$ and set $Q_1=\frac{2e^{ct_1t_3}-e^{ct_2t_3}}{e^c}$, $v=Q_1-Q_1^2$. Here $Q_1$ and $v$ are the expected value and variance of Bernoulli Ber$(q_i(U_i))$ random variable.
\end{enumerate}
Also we write $h(\cdot,\cdot)$ as the relative entropy function defined in \cite{arratia89}.

\subsubsection{Calculating lower bound on K}
We need to find the lower bound $K_{min}$ of $K$ such that with large probability we have at least $Q$ target sequences in the third urn. Equivalently, we want $R \geq Q$. To this end, we assume the cutting process in urn 2 does not break any target sequences and we take everything out from urn 3. Therefore, we only need to make sure $X$ is larger than $Q$ with high probability. In section~\ref{sampling:sec:numerical}, we solved both \eqref{sampling:eq:goal_1} and \eqref{sampling:eq:goal_2} to get different lower bounds for $K$, similarly with different lower bounds on $K$ we will have different lower bounds for downstream quantities like $X$, $U$ etc. 

\subsubsection{Calculating lower bound on R}
Algorithm~\ref{sampling:Algo:01} can be used to calculate $\hat{R}_{low}$ with pre-fixed $n$ and $K$. Please note that we use tail bounds of binomial distribution to approximate that of hyper-geometric distribution in step 1, 2 and step 4. Here step 1 and 2 only requires property 1 and 2 in section~\ref{sampling:sec:hyper_proof}, while for step 4 we also need property 3 and 4, because we need the relations in \eqref{sampling:eq:goal_1} and \eqref{sampling:eq:goal_2} to be true with $W \geq W_{low}$ and $U \leq U_{up}$ as well. For each fixed $n$, the range of $K$ is relatively small, thus for each input $n$ we can simply try all the possible $K$ and calculate the corresponding smallest $R$ (use $R_{low}$ to denote it) that achieves our goal. To make our algorithm more efficient, we can first find the smallest $K$ that can give us a lower tail that is larger than $Q$ (any smaller $K$ will not be feasible, see our supporting codes for details), call this $K_{min}$. For each $K$ from $K_{min}$ to $45n$, we use Algorithm~\ref{sampling:Algo:01} to find $R_{low}$.

\begin{algorithm}[h]
\caption{Computing $\hat{R}_{low}$ from fixed $n$ and $K$}\label{sampling:Algo:01}
\begin{algorithmic}[1]
\State Apply lemma~\ref{sampling:lem:largedeviation} to $B_a \sim Bin(K,\frac{1}{46})$ and $B_b \sim Bin(46n-K,\frac{1}{46})$. Solve the following system:
\begin{align*}
    -log(1-p_1) = K h(\frac{t}{K} + 1-\frac{1}{46}, 1-\frac{1}{46}),
\end{align*}
and set $X_{low_1} = \frac{K}{46} - t$. Similarly we can solve for $X_{low_2}$. Set $X_{low} = \max(X_{low_1}, X_{low_2})$.

\State Now fix $X$ to be $X_{low}$. Solve the following system 
\begin{align*}
    -log(1-p_2) = X_{low} h(\frac{t}{X_{low}}+1-Q_1,1-Q_1),
\end{align*}
and set $W = Q_1*X_{low}-t$.

\State Calculate the $p_3$ lower and upper bounds for $U$ from lemma~\ref{sampling:lem:poisson}.
\State Apply lemma~\ref{sampling:lem:largedeviation} to $B_c \sim Bin(R, \frac{W_{low}}{U_{up}})$ and $B_d \sim Bin(U_{up}-R, \frac{W_{low}}{U_{up}})$, and solve for $R_{low_1}$ from the following system
\begin{align*}
    &rW_{low}/U_{up}-t=Q, \\
    &-log(1-p_0)=r h(\frac{t}{4}+1-\frac{W_{low}}{U_{up}},1-\frac{W_{low}}{U_{up}}),
\end{align*}
\State Set $\hat{R}_{low} = r$ and output $(n,K,\hat{R}_{low})$. 
\end{algorithmic}
\end{algorithm} 

\subsection{Lemmas}
To make this paper self-contained, we list the lemmas used in our calculation in this section. Detailed proof can be found in relevant references. 
\begin{lem}\label{sampling:lem:poisson}(Bounds on Poisson distribution. See \cite{michael13}) For $U$ defined in section~\ref{sampling:sec:model} and $p \in (0,1)$, we have
\begin{align}
& U \leq cK+K+\Phi^{-1}(p)\sqrt{cK}+\frac{\Phi^{-1}(p)^2}{6}, \\
& U \geq cK+K-\sqrt{-2cK ln(1-p)},
\end{align}
all with probability at least $p$.
\end{lem}

\begin{lem}\label{sampling:lem:largedeviation}(Large deviation bound on binomial distribution. See \cite{arratia89}) Let $X = \sum_{i=1}^m X_i$, here $\lbrace X_i \rbrace_i^{m}$ are i.i.d. Bernoulli trials with probability of success equal to $p$. Assume $p < a < 1$ for constant $a$. Use $h$ to denote the relative entropy (defined in \cite{arratia89})  between $a$ and $p$. Then 
\begin{equation*}
    \mathbb{P}\left[ X \geq a m \right] \leq e^{-mh}.
\end{equation*}
This bound is relatively tighter than Chernoff bounds with small $p$.
\end{lem}

{
\small
\bibliographystyle{plainnat}
\bibliography{wileyNJD-APA}

\begin{thebibliography}{6}
\providecommand{\natexlab}[1]{#1}
\providecommand{\url}[1]{\texttt{#1}}
\expandafter\ifx\csname urlstyle\endcsname\relax
  \providecommand{\doi}[1]{doi: #1}\else
  \providecommand{\doi}{doi: \begingroup \urlstyle{rm}\Url}\fi

\bibitem[Arratia and Gordon(1989)]{arratia89}
Richard Arratia and Louis Gordon.
\newblock Tutorial on large deviations for the binomial distribution.
\newblock \emph{Bulletin of mathematical biology}, 51\penalty0 (1):\penalty0
  125--131, 1989.

\bibitem[Audano et~al.(2019)Audano, Sulovari, Graves-Lindsay, Cantsilieris,
  Sorensen, Welch, Dougherty, Nelson, Shah, Dutcher, et~al.]{audano19}
Peter~A Audano, Arvis Sulovari, Tina~A Graves-Lindsay, Stuart Cantsilieris,
  Melanie Sorensen, AnneMarie~E Welch, Max~L Dougherty, Bradley~J Nelson,
  Ankeeta Shah, Susan~K Dutcher, et~al.
\newblock Characterizing the major structural variant alleles of the human
  genome.
\newblock \emph{Cell}, 176\penalty0 (3):\penalty0 663--675, 2019.

\bibitem[Chaisson et~al.(2015)Chaisson, Wilson, and Eichler]{chaisson15}
Mark~JP Chaisson, Richard~K Wilson, and Evan~E Eichler.
\newblock Genetic variation and the de novo assembly of human genomes.
\newblock \emph{Nature Reviews Genetics}, 16\penalty0 (11):\penalty0 627, 2015.

\bibitem[Huddleston and Eichler(2016)]{huddleston16}
John Huddleston and Evan~E Eichler.
\newblock An incomplete understanding of human genetic variation.
\newblock \emph{Genetics}, 202\penalty0 (4):\penalty0 1251--1254, 2016.

\bibitem[Hurles et~al.(2008)Hurles, Dermitzakis, and Tyler-Smith]{hurles08}
Matthew~E Hurles, Emmanouil~T Dermitzakis, and Chris Tyler-Smith.
\newblock The functional impact of structural variation in humans.
\newblock \emph{Trends in Genetics}, 24\penalty0 (5):\penalty0 238--245, 2008.

\bibitem[Short(2013)]{michael13}
Michael Short.
\newblock Improved inequalities for the poisson and binomial distribution and
  upper tail quantile functions.
\newblock \emph{ISRN Probability and Statistics}, 2013.

\end{thebibliography}
}
\end{document}